# Transient and Periodic Steady-State Characteristics of the Local Heat Transfer Measurement by Thermal Perturbation with Gaussian Power Density Distribution

Zhongyuan Shi [a,b,*,†] Tao Dong [c,*,†] Zhaochu Yang [a,b]

a. Chongqing Municipality Key Laboratory of Micro-nano Systems and Intelligent Transduction

   Chongqing Technology and Business University, 19 Xuefu Ave., Nan'an District, Chongqing, China

b. Chongqing Key Laboratory of Colleges and Universities on Micro-nano Systems Technology and Smart Transduction, Collaborative Innovation Center of Micro-Nano Transduction and Intelligent Ecological Internet of Things (IoT)

   Chongqing Technology and Business University, 19 Xuefu Ave., Nan'an District, Chongqing, China

c. Department of Microsystems

   Faculty of Technology, Natural Sciences and Maritime Sciences, University of South-Eastern Norway (USN), Raveien 215, 3184 Borre, Norway

\* The corresponding authors' email address: Tao.Dong@usn.no

† The two authors, Zhongyuan Shi (zhongyuan.shi@ctbu.edu.cn) and Tao Dong, have the same contribution to this study.

**Abstract**

The local heat transfer coefficient measurement with temperature oscillation induced by periodic thermal perturbation – usually via a Gaussian laser beam, was investigated for the impact of the *spikiness* (i.e., the standard deviation) elaborated in comparison with the analytical model for dimensional analysis. The statistically more robust technique that relies on the linearity of the spatial phase distribution of the test point array was favored when the target Biot number approaches unity in terms of its order of magnitude. The preferred upper limit for thermographic scanning was discussed as the simplification of later data processing is concerned. Nonetheless, the time elapsed for an acceptable periodic steady state, which in principle leans to the higher end of the target Biot number spectrum in a log scale, indicates the benefit from the time series of pointwise temperature measurement – as in the conventional single-blow testing, where the effect of *spikiness*, as well as that of the location of individual test point, holds. Note that the vicinity as the target Biot number approaches unity was again observed with higher preference.

**Nomenclature**

| | |
|---|---|
| $Bi$ | Biot number |
| $h$ | heat transfer coefficient, W/(m²K) |
| $i$ | imaginary unit |
| $j$ | node indice along the $\tilde{r}$ axis, j=1, 2, … |
| $k$ | node indice along the $\tilde{z}$ axis, k=1,2, …, N |
| $q$ | heat flux, W/m² |
| $\tilde{q}$ | heat flux, dimensionless |
| $r$ | coordinate $r$ in the 2D cylindrical system, m |
| $\tilde{r}$ | coordinate $\tilde{r}$ in the 2D cylindrical system, dimensionless |
| $T$ | temperature, K |
| $t$ | time, s |
| $T_\infty$ | reference temperature, K |
| $z$ | coordinate z in the 2D cylindrical system, m |
| $\tilde{z}$ | coordinate z in the 2D cylindrical system, dimensionless |

**Greek symbols**

| | |
|---|---|
| $\alpha$ | thermal diffusivity, m²/s |
| $\beta$ | slope of the phase line, see Fig. 5 |
| $\delta$ | thickness, m |
| $\tilde{\delta}$ | thickness, dimensionless |
| $\Delta\phi$ | phase, rad |
| $\eta$ | parameter to indicate different wall treatments, see Eq. (7) |
| $\theta$ | temperature, dimensionless |
| $\kappa$ | the temperature rise at the end of the first cycle in the time domain, see Fig. 8 |
| $\lambda$ | thermal conductivity, W/(m·K) |
| $\sigma$ | standard deviation |
| $\tau$ | time, dimensionless |
| $\omega$ | oscillation frequency, Hz |

**Subscripts**

| | |
|---|---|
| m1c | mark at the end of the first cycle of the phase line, see Fig. 5 |
| PSS | periodic steady state |
| ref | reference |
| tp | test point |

## 1. Introduction

Efficient heat transfer coefficient measurement is indispensable to iterative design and daily operation in numerous thermal processes, wherever experimental validation, evaluation, or diagnostics of heat transfer devices are involved. Based on inverse problem settings [1]-[3], transient techniques, particularly those featured with local measurement and contactless-ness, are gaining interest from investigators worldwide, while the convention can be traced well back to Nusselt's time [4]-[6] when the outlet temperature response of the porous medium, subject to an abrupt temperature change of the inlet fluid, was investigated.

Furnas [7] developed the model for acquiring the heat transfer coefficient of compact units. The method is now much more widespread with the tag *single-blow testing*. Kohlmayr [8]-[11] perceived the initial rise with the maximum slope in the logged temperature sequence at the outlet as an important indicator, analytically, in response to arbitrary temperature variation at the inlet. Liang and Yang [12] also presented the analytical solution to an exponential boundary. Later development [13]-[23] on the effect of axial thermal diffusion, lateral thermal resistance, fin dynamics, and so forth further enriched the literature.

In parallel to the development of single-blow testing, Hausen and his followers [24][25] initiated a contemporary exploration of the value of periodic temperature variation in determining the heat transfer coefficient. Kast [26], Matulla and Orlicek [27] focused on sinusoidal oscillations at the inlet and outlet of the test section. Inspired by the preceding work, Roetzel [28] believes that the amplitude and phase of the temperature oscillation at any specific location of the conjugated system can be employed for the measurement with a reference, the waveform being irrelevant. The follow-up implementation on the local heat transfer coefficient measurement exploited the thermal wave from a flicker laser perturbation on the dry side of the heat transfer surface (i.e., the wet side), where convection dominates. The temperature field was then recorded in sequence for the spatial variation of phase [29][30] or its mean value relative to that of the thermal perturbation [31]-[36] to be obtained and related to the heat transfer coefficient on the fluid side analytically. A similar one-dimensional analytical solution from Turnbull and Oosthuizen [37] showed that the square wave also works with the phase delay method. A turbine blade experiment later validated the numerical results [38]. Freund and his colleagues [39]-[41] extended the boundary to real-world applications with a thermal-electric model, which facilitates the phase-to-heat-transfer-coefficient process based on a periodic steady-state (PSS) analysis. The iterative method was expected to contour the surface variation of the heat transfer coefficient with a single perturbation heat source that again covers the target from the dry side. Starting from the one-dimensional cylindrical system with PSS, Leblay et al. [42]-[44] presented the reliability of the in-tube heat transfer coefficient measurement with a dual-frequency sinusoidal thermal excitation at the exterior wall, where the corresponding amplitude of the temperature oscillation was compared to the analytical solution.

In the present study, the PSS analysis pertains to the correlation between the phase of the induced temperature oscillation and the heat transfer coefficient. The relationship has been experimentally validated by the authors [35]. The dimensional analysis is aimed at exploring the design space for the measurement. The standard deviation of the widely used Gaussian laser beam [40][45], used to quantify how asymptotic the scenario is referring to the classical analytical model in the one-dimensional configuration, was found to affect the measurement principle substantially in the present work. The result from dimensional analysis indicates whether the maximum slope variant from single-blow testing or the outcome from the PSS thermal-electric analogy should be taken as a guideline for data processing, depending on the location preference of the test points, as well as the estimated ratio of heat transfer density from convection to that from thermal diffusion.

## 2. Model Description

As shown in Fig. 1, the perturbation thermal input from above comes with a normalized Gaussian distribution in regard to its power density. The axis of the cylindrical system separates the right half of the solid domain, of which the bottom side is subject to a uniform convectional surrounding.

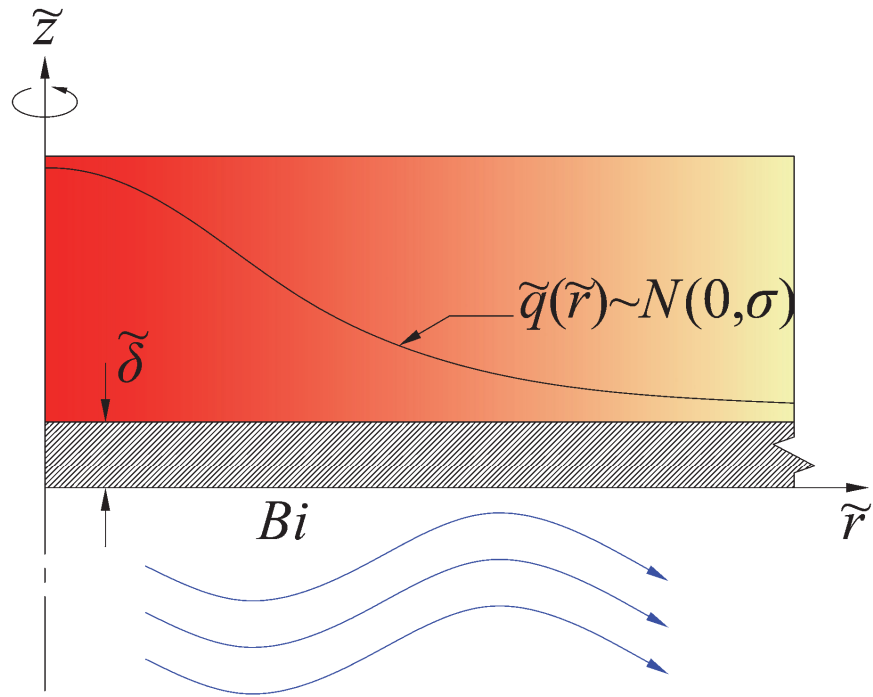

Fig. 1 Measurement principle in a 2D cylindrical system

For dimensional analysis,

$$\begin{cases} \dfrac{\partial \theta}{\partial \tau} = \dfrac{\partial^2 \theta}{\partial \tilde{r}^2} + \dfrac{1}{\tilde{r}} \dfrac{\partial \theta}{\partial \tilde{r}} + \dfrac{\partial^2 \theta}{\partial \tilde{z}^2} \\ \left. \dfrac{\partial \theta}{\partial \tilde{z}} \right|_{\tilde{r}=0\sim\infty, \tilde{z}=1} = \dfrac{(1 + \text{sgn} \sin 2\pi\tau) e^{-\tilde{r}^2/(2\sigma^2)}}{2\sigma\sqrt{2\pi}} \\ \left. \dfrac{\partial \theta}{\partial \tilde{r}} \right|_{\tilde{r}=0, \tilde{z}=0\sim1} = 0 \\ \left. \dfrac{\partial \theta}{\partial \tilde{z}} \right|_{\tilde{r}=0\sim\infty, \tilde{z}=0} = Bi\theta \end{cases}, \qquad (1)$$

where

$$\begin{cases} \theta = \dfrac{\lambda(T - T_\infty)}{q_{ref}} \sqrt{\dfrac{\omega}{\alpha}} \\ \tilde{r} = r\sqrt{\dfrac{\omega}{\alpha}}, \tilde{\delta} = 1 \\ \tilde{z} = z\sqrt{\dfrac{\omega}{\alpha}} \\ \tau = \omega t \\ Bi = \dfrac{h}{k}\sqrt{\dfrac{\alpha}{\omega}} \end{cases}. \qquad (2)$$

The dimensionless thickness $\tilde{\delta}$ was set to unity as the thermal perturbation frequency $\omega$ is determined in the meanwhile. The Biot number represents the to-be-measured heat transfer coefficient as referring to the thermal conductivity of the wall material. The definition of the characteristic length $\sqrt{\alpha/\omega}$ could be regarded as a measure of thermal diffusion through the wall material within a perturbation period [30].

The method is aimed at the extraction of the heat transfer coefficient from the measured phase array of the test points aligned radially away from the center of the thermal perturbation (presumably a circular laser spot, for instance), the PSS analysis [46] turns the above-listed governing equations as well as the associated boundary conditions into

$$\begin{cases} i\theta\tilde{r} = \dfrac{\partial}{\partial\tilde{r}}\left(\tilde{r}\dfrac{\partial\theta}{\partial\tilde{r}}\right) + \tilde{r}\dfrac{\partial}{\partial\tilde{z}}\left(\dfrac{\partial\theta}{\partial\tilde{z}}\right) \\ \left.\dfrac{\partial\theta}{\partial\tilde{z}}\right|_{\tilde{r}=0\sim\infty,\tilde{z}=1} = -i\dfrac{e^{-\tilde{r}^2/(2\sigma^2)}}{\sigma\sqrt{2\pi}} \\ \left.\dfrac{\partial\theta}{\partial\tilde{r}}\right|_{\tilde{r}=0,\tilde{z}=0\sim 1} = 0 \\ \left.\dfrac{\partial\theta}{\partial\tilde{z}}\right|_{\tilde{r}=0\sim\infty,\tilde{z}=0} = Bi\theta \end{cases} \quad (3)$$

The waveform does not make a difference in PSS while serving as an amenity for the time domain [28][40].

ANSYS Workbench [47] was employed for the numerical solution to Eqs. (1)-(2). The MATLAB code from Freund's dissertation [40] was adapted with the generalized minimal residual method (GMRES) [48] to solve Eq. (3) in the discretized form, as elaborated in Eqs. (4)-(8) in the following sections (2.1 & 2.2). The grid independence, as well as the far field condition, i.e., where $\tilde{r} \to \infty$ and $\tilde{z} = 0\sim 1$, was checked regarding the concerned features in Section 3.

2.1 Inner Node

Integrating over the inner control volume (CV$_{k,j}$) yields

$$i\theta_{k,j}\tilde{r}_{k,j}\delta\tilde{r}\delta\tilde{z} = \tilde{r}_{k,j+1/2}\dfrac{\theta_{k,j+1}}{\delta\tilde{r}}\delta\tilde{z} - 2\tilde{r}_{k,j}\dfrac{\theta_{k,j}}{\delta\tilde{r}}\delta\tilde{z} + \tilde{r}_{k,j-\frac{1}{2}}\dfrac{\theta_{k,j-1}}{\delta\tilde{r}}\delta\tilde{z} \\ + \tilde{r}_{k,j}\dfrac{\theta_{k+1,j}}{\delta\tilde{z}}\delta\tilde{r} - 2\tilde{r}_{k,j}\dfrac{\theta_{k,j}}{\delta\tilde{z}}\delta\tilde{r} + \tilde{r}_{k,j}\dfrac{\theta_{k-1,j}}{\delta\tilde{z}}\delta\tilde{r}. \quad (4)$$

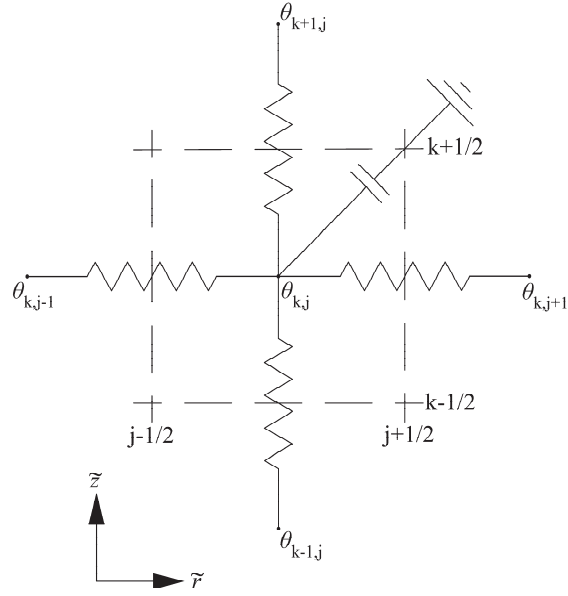

Fig. 2 A schematic representation of the control volume ($CV_{k,j}$) for PSS analysis

## 2.2 Wall Nodes

For the wall nodes with perturbation heat flux,

$$-i\frac{e^{-\frac{\tilde{r}_j^2}{(2\sigma^2)}}}{\sigma\sqrt{2\pi}} = \eta\frac{2(\theta_{N,j} - \theta_{N-1,j})}{\delta\tilde{z}} + (\eta - 1)\frac{\theta_{N-2,j} - \theta_{N-1,j}}{\delta\tilde{z}}, \qquad (5)$$

where [49]

$$\eta = \begin{cases} \frac{4}{3}, higher-order\ treatment \\ 1, lower-order\ treatment \end{cases}. \qquad (6)$$

The infinitesimal thickness of the control volume represented by the wall node (N,j) holds no heat capacity while the one next to it, i.e., node (N-1,j) in this case, does. Kirchhoff's current law applied to node (N-1,j) differs with the half distance to the wall node compared to other inner nodes. The adjacent nodes, i.e., (N-1,j-1) and (N-1,j+1), are neglected as the equation holds valid for the wall node (N,j) exclusively.

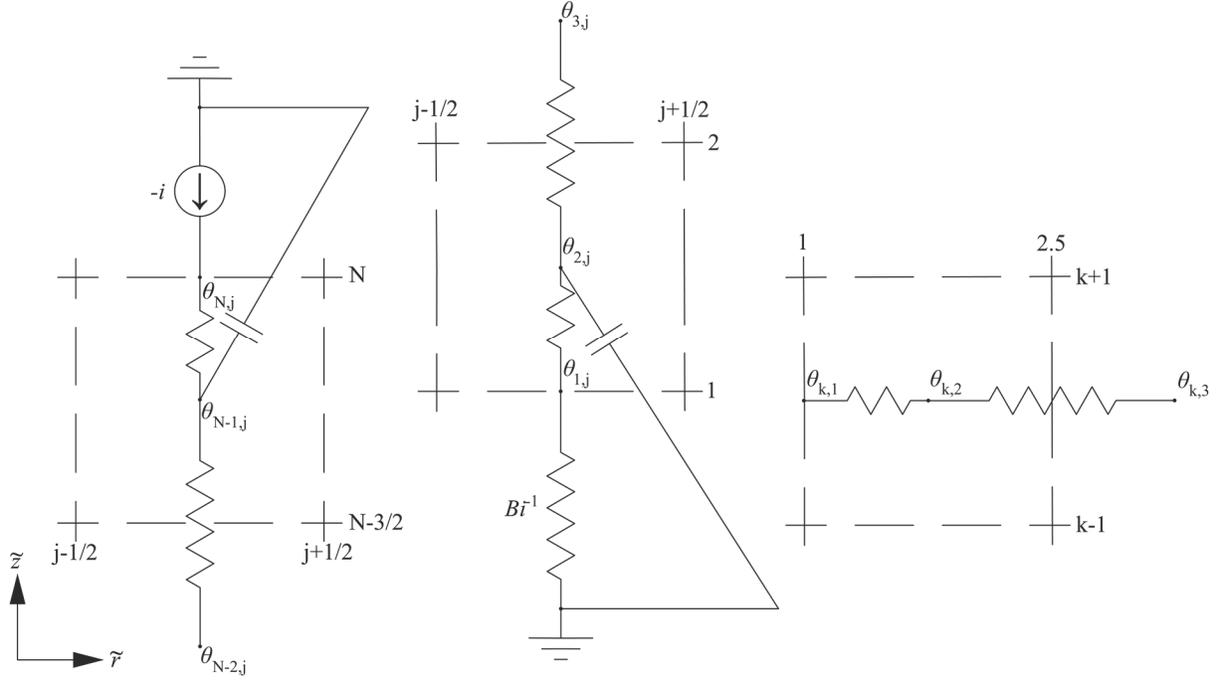

Fig. 3 A schematic representation of the wall node $\theta_{N,j}$ with perturbation heat flux, wall node $\theta_{1,j}$ with convection and the wall node $\theta_{k,1}$ with symmetry in the PSS analysis

Similarly, for the wall nodes with convection

$$-Bi\theta_{1,j} = \eta\frac{2(\theta_{1,j} - \theta_{2,j})}{\delta\tilde{z}} + (\eta - 1)\frac{\theta_{3,j} - \theta_{2,j}}{\delta\tilde{z}}, \tag{7}$$

and for the wall nodes with symmetry

$$0 = \eta\frac{2(\theta_{k,1} - \theta_{k,2})}{\delta\tilde{r}} + (\eta - 1)\frac{\theta_{k,3} - \theta_{k,2}}{\delta\tilde{r}}. \tag{8}$$

## 3. Results and Discussion

The impact from the *spikiness* (interchangeably the standard deviation, $\sigma$) of the perturbation laser spot with Gaussian-distributed heat flux is shown in Fig. 4 (a)-(d), which is presented in terms of the temperature oscillation phase $\Delta\phi$ versus the radial coordinate $\tilde{r}_{tp}$ of the test points located where the laser beam meets the wall. *Spikier* distribution (lower $\sigma$) brings higher similarity to the point-source solution in the one-dimensional configuration. The linearity between $\Delta\phi$ and $\tilde{r}_{tp}$ can be identified toward the trailing edge ($\tilde{r}_{tp} \to 2$) in the best scenario ($\sigma=0.1$), showing the asymptotic validity compared to the analytical solution. However, the initial progression of curvature follows an inverse trend, wherein the slope steadily increases towards a stable value. Decreasing $\sigma$ seems equivalent to zooming in to the onset. The defining feature of Roetzel's method [30] loses its validity as the slope difference diminishes with a more *flattened* distribution, while the variation of $\Delta\phi$ is getting less related to $\tilde{r}_{tp}$. The analytical solution with the modified boundary condition provides a slightly higher resolution for the measurement as Bi increases and deviates from unity, despite the better agreement with the numerical results for the cases with Bi on the order of unity or less.

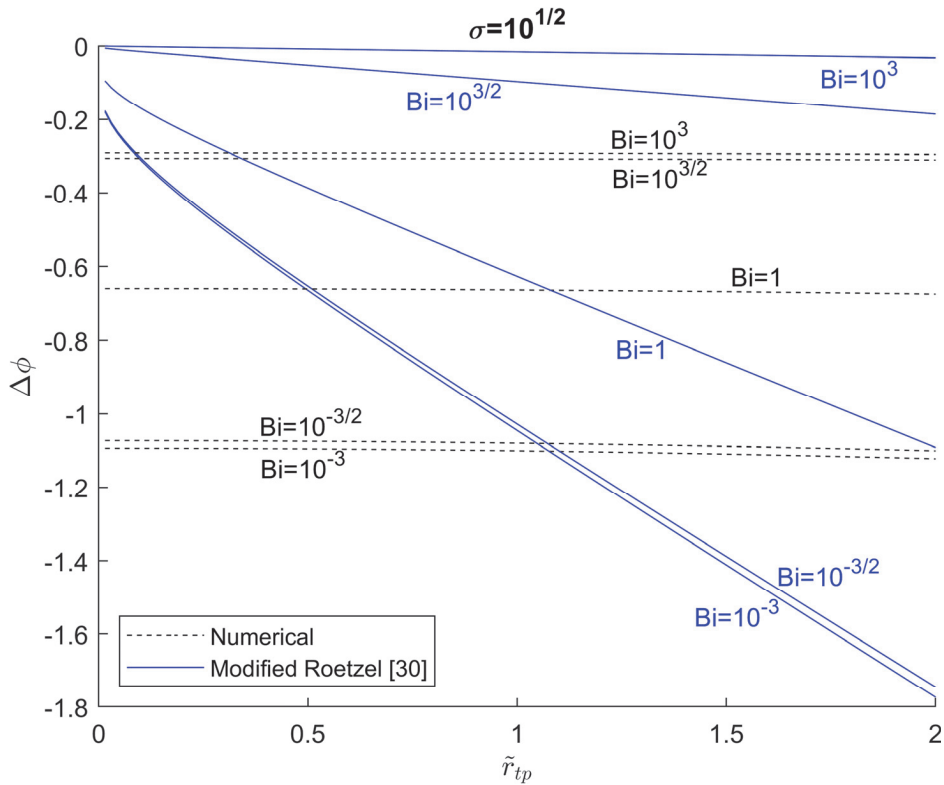

(a)

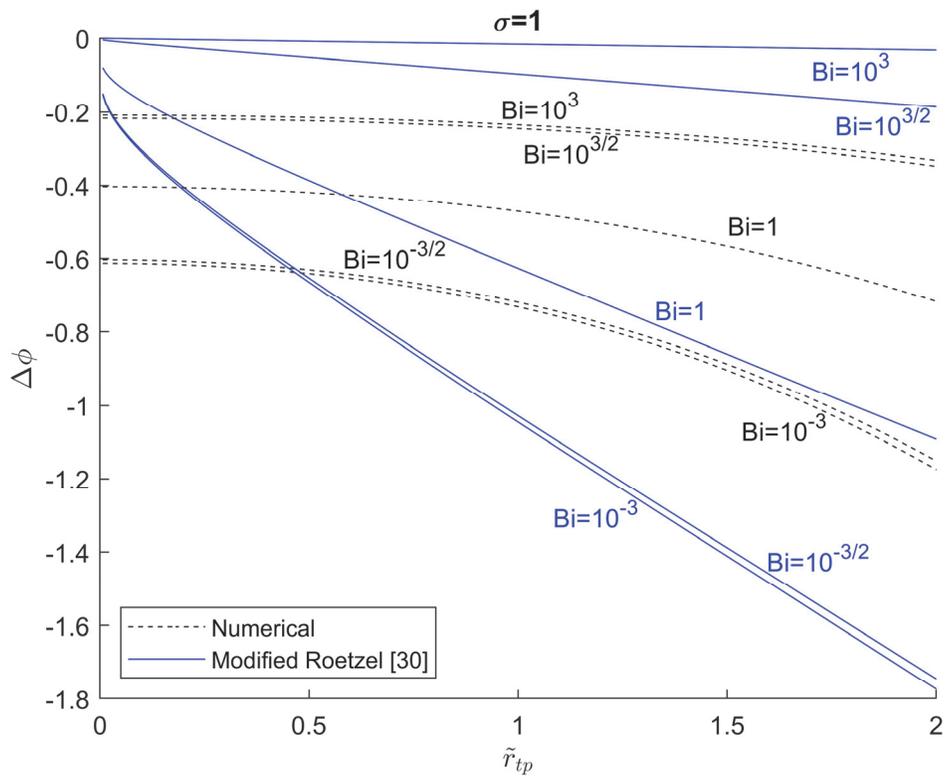

(b)

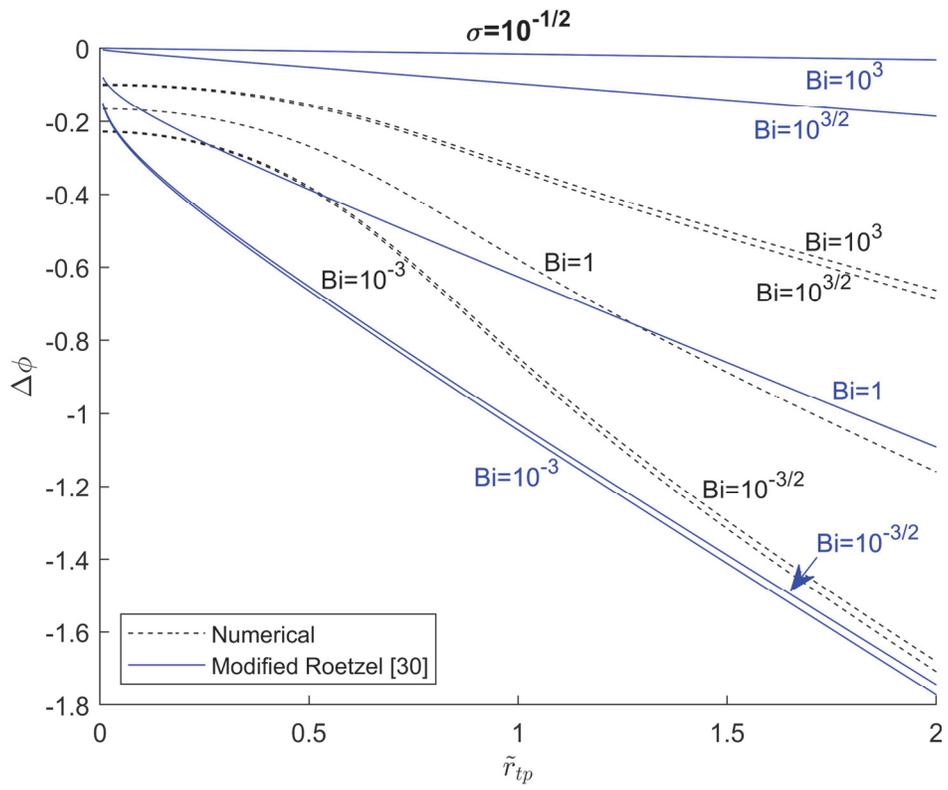

(c)

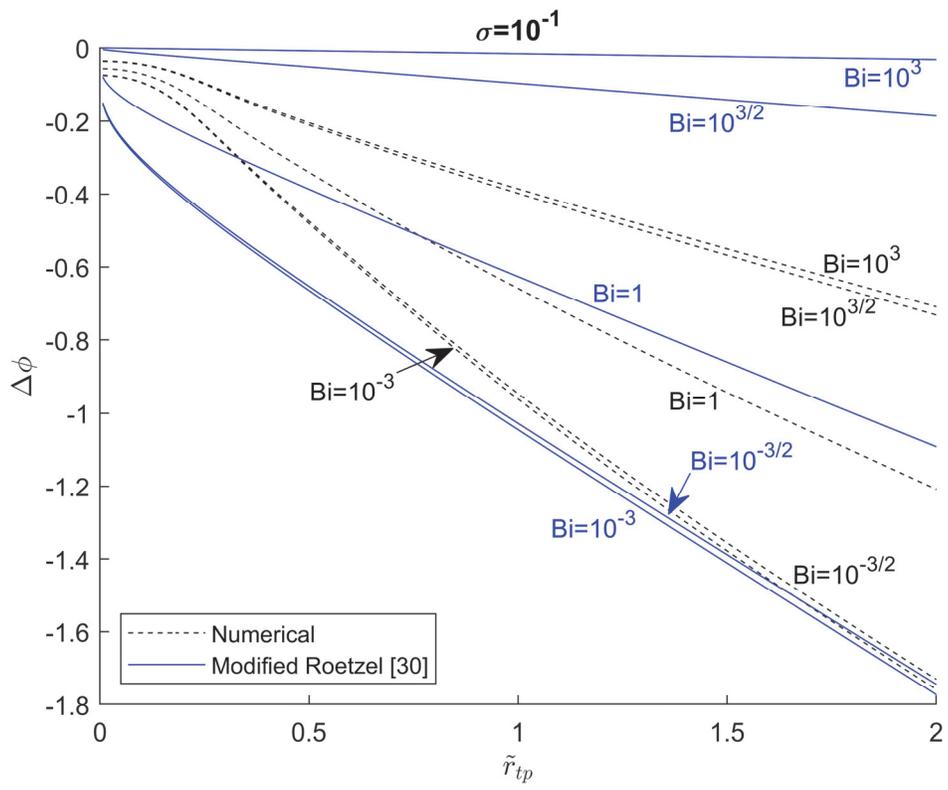

(d)

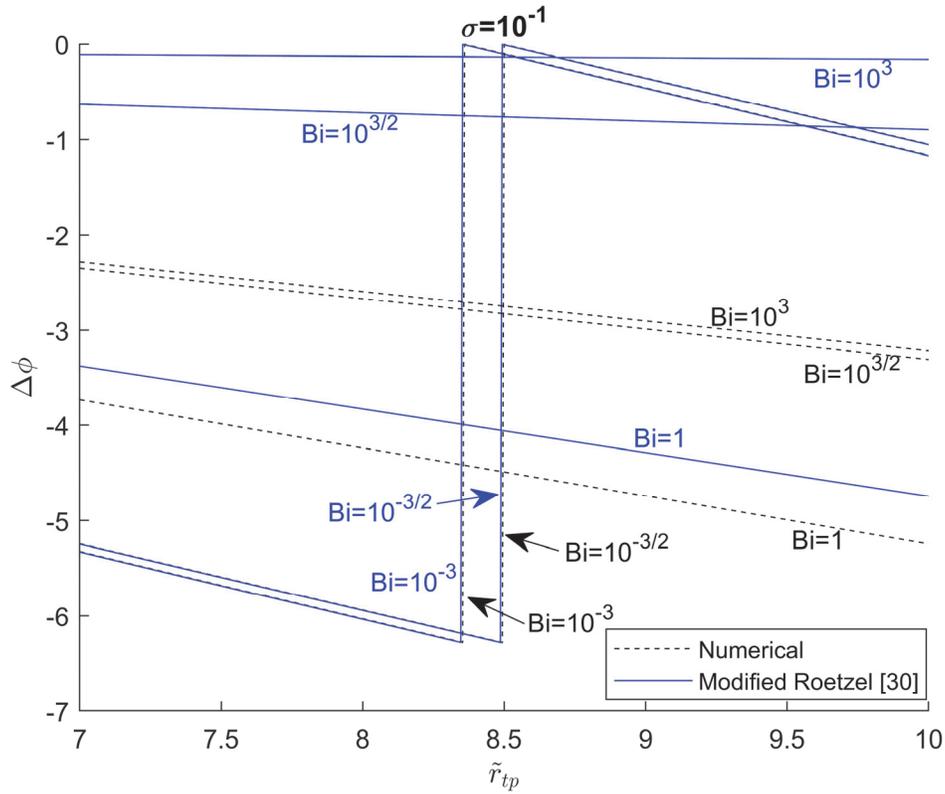

(e)

Fig. 4 (a)~(d) The impact of *spikiness* ($\sigma$) on the phase-to-Bi relationship. The boundary condition in Roetzel's model was replaced with a point source of periodic heat flux. A better agreement was found with the numerical results, particularly where the phase gets re-initialized as in (e) - a prominent feature for Bessel and related functions.

In Fig. 5, the blue curve with triangle symbols marks the slopes of the phase lines as in Fig. 4 (a)~(d). While the general trend favors the case as Bi is instead approaching unity where the highest resolution or the lowest sensitivity of the measurement takes place, as is indicated in all the cases, be it numerical or analytical. The performance-related parameter reaches its maximum or minimum when the covariant approaches unity, which is hereinafter referred to as the Specific Feature of Unity (SFU). It holds in many other dimensional analyses, showing the significance of rescaling in data processing [50], as is the case in Wandelt's method [32], and later in the present work as well. The *spikiness* has negligible influence on the slope ($\beta$) of the phase lines, especially as those in Fig. 4 (c)~(d) with more pertinence to the section away from the center of the imposed heat flux, where the linearity between $\Delta\phi$ and Bi is more prominent. Following the trend in Fig. 4, the condition to acquire thermal diffusivity of the wall material (Bi=0) [30] was set in Fig. 5 with a red curve and circular marks to depict the effect of *spikiness* on the locations ($\tilde{r}_{tp,m1c}$) of the test points, where the phase gets re-initialized, and the injective mapping from $\Delta\phi$ to Bi is compromised, owing to the periodic nature of Bessel and related functions. The locations ($\tilde{r}_{tp,m1c}$), being the distance from the center of the imposed perturbation heat flux and more than doubled as the *spikiness* increases by one and a half orders of magnitude, therefore, define the upper bounds for measurement, a notable feature as the array size of infrared detectors is concerned [40].

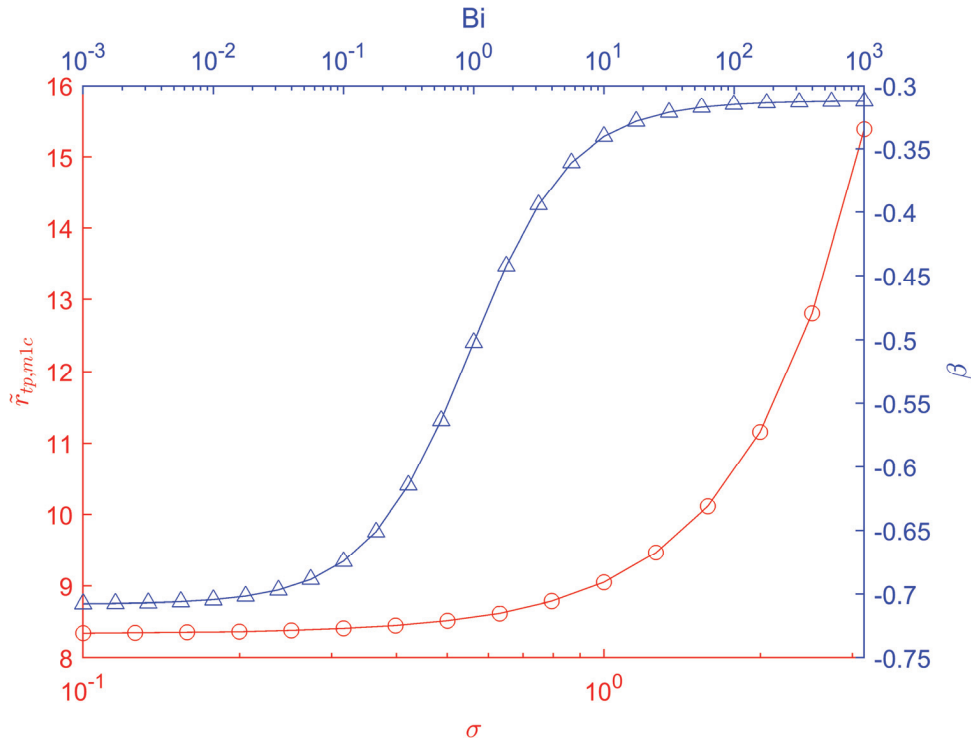

Fig. 5 Variation of the phase line's slope $\beta$ [30] with Bi, marked with the blue triangle, and that of $\tilde{r}_{tp,m1c}$, which marks the end of the 1$^{st}$ cycle of the phase line, with the *spikiness* ($\sigma$)

Insufficient cooling leads to thermal build-up and a more extended period of transient response at the test point, as the Biot number decreases for the same *spikiness* of the perturbing laser input, as is identifiable in Fig. 6. Note that this is not rendering any possibility of boosting the Biot number by lowering the perturbation frequency, considering the cost from the real-world time to wait for an acceptable PSS according to the dimensional setting in Eq. 2, which is aimed at a more standardized procedure for data processing starting from the appropriate choice of the length scale $\sqrt{\alpha/\omega} = \delta$.

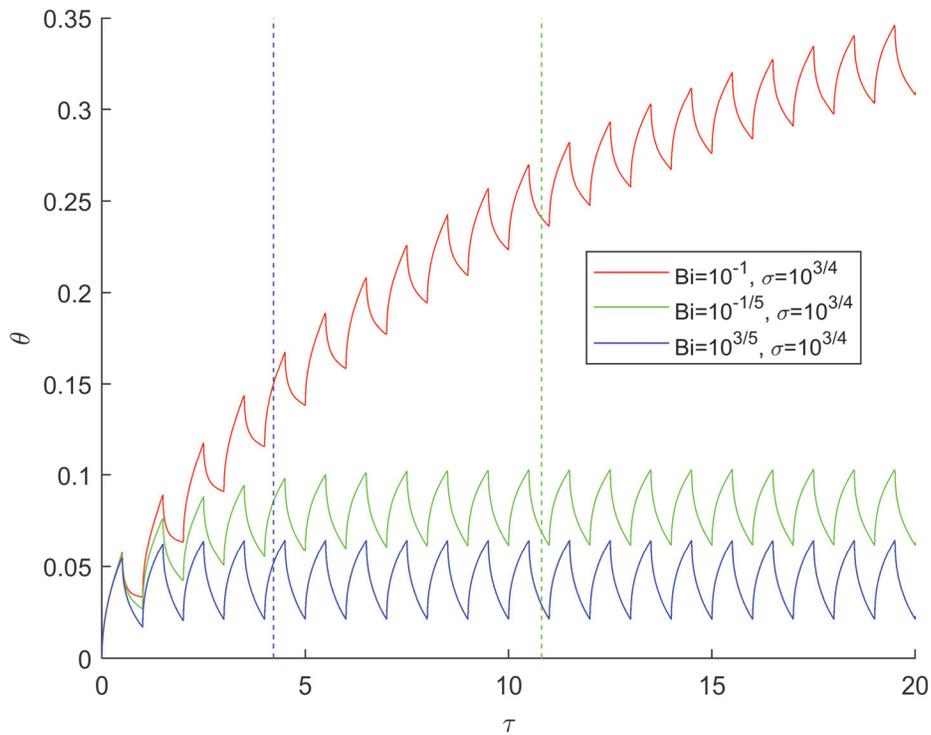

Fig. 6 Typical temperature rises in the time domain with different convections (Bi), dashed lines with different colors mark their respective time elapsed before PSS ($\tau_{PSS}$ in Fig. 7, defined as the temperature difference $\theta(\tau) - \theta(\tau - 1)$ is within 1% of the average amplitude [40]). For the case with Bi=$10^{-1}$, the dashed line in red goes far beyond the upper limit.

Fig. 7 shows the impacts on the dimensionless time before an acceptable PSS, which stabilizes itself toward the higher end of the Bi spectrum, from the *spikiness* and the location of the test point.

The test point with a unit length away from the laser center performs the best in general, while increasing the *spikiness* was proven detrimental as opposed to the case with the test point located ten times the distance where higher *spikiness* is expected. The trend lines tend to collapse or converge into a single line as the test point is further approaching the center of the laser spot, with inferior overall performance compared to the case with the test point located a unit length away.

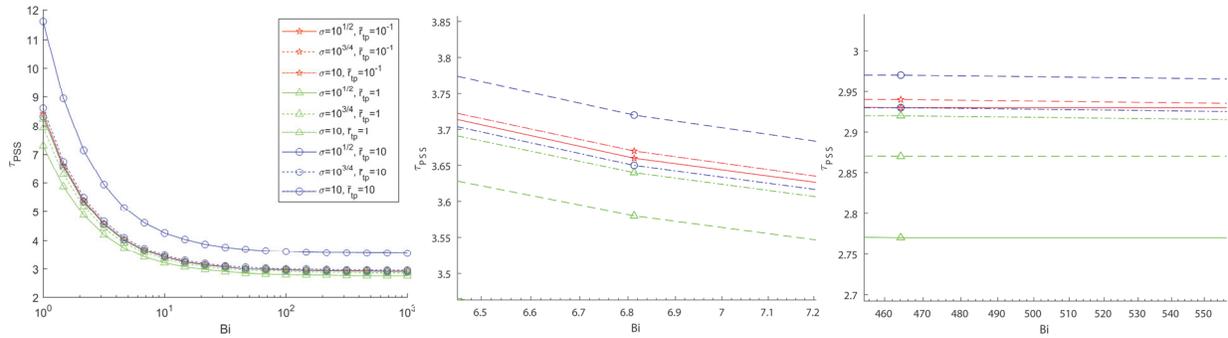

Fig. 7 Trendlines of the time elapsed before PSS ($\tau_{PSS}$) vs. Bi, with locally enlarged views to better illustrate the region where points and/or curves cluster

Noteworthily, Fig. 6 is a reminder of the classic maximum slope method in single-blow testing [21], a close relative in the family of transient techniques. The temperature rise at the end of the 1$^{st}$ cycle, denoted as $\kappa$, represents the slope in Fig. 8. The majority of its variations occur approximately as Bi increases from $10^{-1}$ to 10 - another exemplification of the SFU. Again, the *spikiness* kicks in and determines the resolution of the measurement. The range of the slope variation differs by an order of magnitude, with the region of favor, as shown in Fig. 9, in the shape of a heated katana tip, centered slightly higher than unity as the *spikiness* increases, while the test point is constrained, roughly, within a unit length from the center of the laser spot.

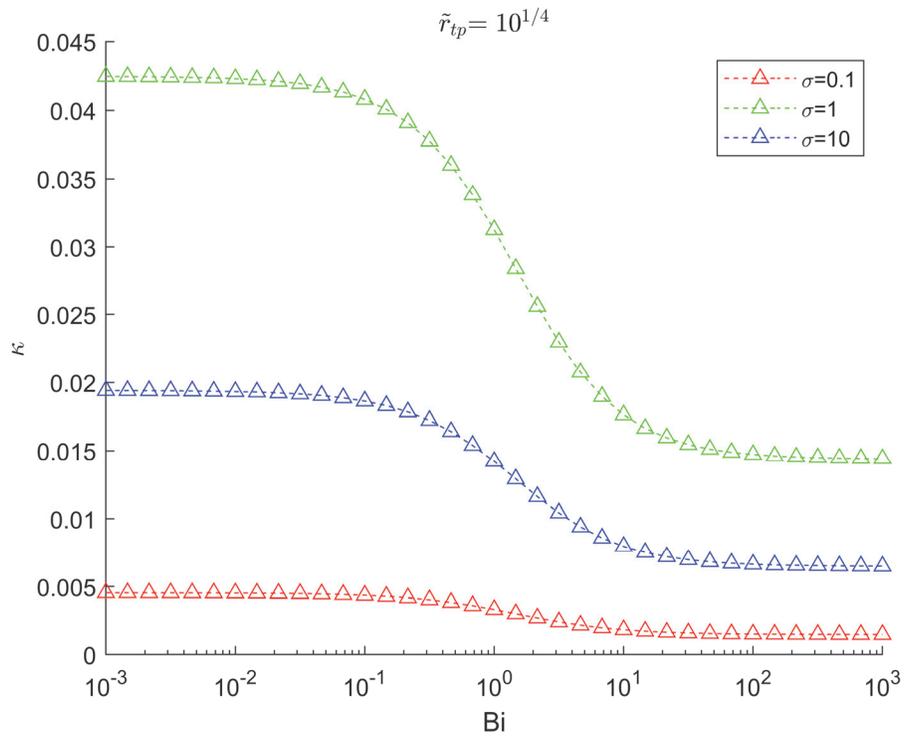

Fig. 8 The impact on $\kappa$, defined as the temperature rise at the end of the 1st cycle in the time domain, from Bi

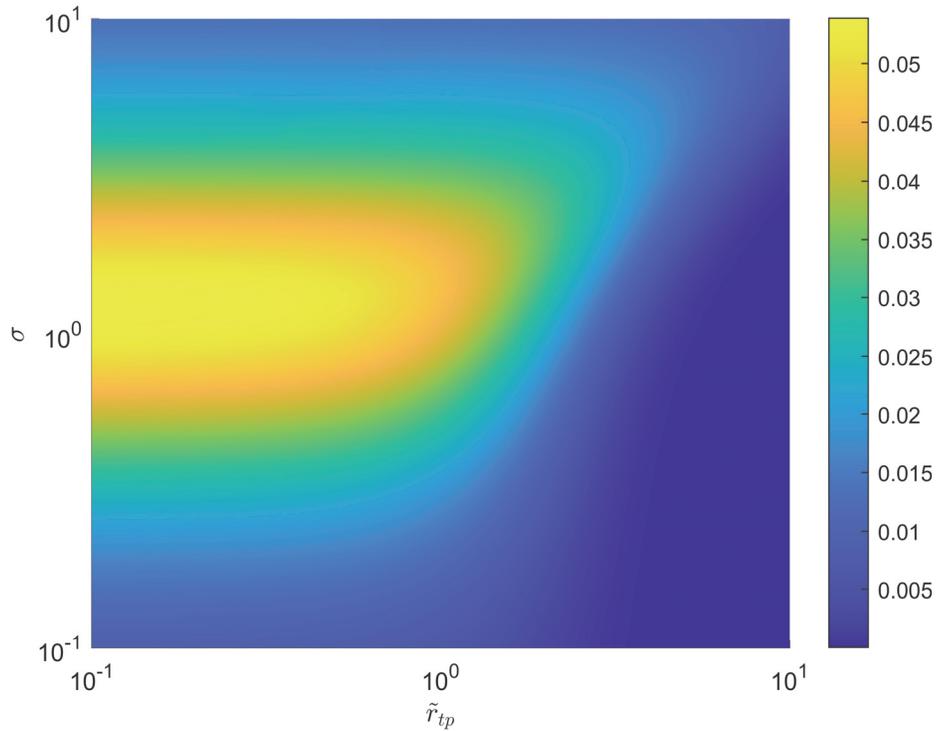

Fig. 9 Contours for the corresponding range of $\kappa$ with the variation of $\tilde{r}_{tp}$ and $\sigma$, Bi spanning from $10^{-3}$ to $10^3$

## 4. Concluding Remarks

For local heat transfer coefficient measurement with temperature oscillation, the effect of the standard deviation in the widely used source of thermal perturbation – the Gaussian laser beam was illustrated as the asymptoticity to the analytical solution and the scope of the captured sequence of thermal images for data reduction. The range of exploration pertaining to the relevant dimensional groups or parameters was determined based on their respective influence on the measurement sensitivity. Particularly, the target Biot number ranges from $10^{-3}$ to $10^3$, compared to the standard deviation between $10^{-1}$ and 10. The time elapsed for an acceptable PSS tends to be negligible to the higher end of the Biot number spectrum (approximately from 10 to $10^3$), as the non-periodic transient response and the follow-up detrending process is considered a primary source of measurement uncertainty. In contrast, a step back to the spirit of the maximum slope method in the time domain may call for retrieving the standard deviation as a parameter, in addition to the location for *pointwise* temperature measurement, as the Biot number is more within the vicinity of unity.

**Acknowledgment**

The financial support from the National Key Research and Development Program of China (No. 2020YFB2009100), Chongqing Municipal Bureau of Science and Technology (No. cstc2021jcyj-msxmX0004), and Chongqing Municipal Education Commission (No. KJQN202200809) is hereby acknowledged.

The present work is funded by the National Natural Science Foundation of China (No. 11702045).

# A Supplementary Perspective to "Transient and periodic steady-state characteristics of the local heat transfer measurement by thermal perturbation with Gaussian power density distribution" with Comments

Zhongyuan Shi[a,†,*] Tao Dong[b,c,a,†,*] Zhaochu Yang[a]

a. Chongqing Key Laboratory of Micro-Nano Transduction and Intelligent Systems, Collaborative Innovation Center on Micro-Nano Transduction and Intelligent Eco-Internet of Things, Chongqing Key Laboratory of Colleges and Universities on Micro-Nano Systems Technology and Smart Transducing, National Research Base of Intelligent Manufacturing Service

   Chongqing Technology and Business University, 19 Xuefu Ave., Nan'an, Chongqing, 400067, China

b. X Multidisciplinary Research Institute, School of Instrument Science and Technology

   Xi'an Jiaotong University, Xi'an, 710049, China

c. State Key Laboratory for Manufacturing Systems Engineering
   Xi'an Jiaotong University, Xi'an, 710049, China

†The two authors, Zhongyuan Shi (zhongyuan.shi@ctbu.edu.cn) and Tao Dong (tao.dong@xjtu.edu.cn), share the same contribution to the present work.

*The corresponding authors' email address: zhongyuan.shi@ctbu.edu.cn, tao.dong@xjtu.edu.cn

**Abstract**
A supplementary perspective was provided to the concerns, including noise tolerance, sampling rate, test duration, the spikiness of the imposed heat flux, and the accuracy-related parameters, in the measurement of local Biot number with thermal perturbation. The optimization was implemented with a Gaussian process surrogate model for data processing, within the specified parametric range of interest. The two commonly employed temporal modes of the imposed heat flux were compared with counterintuitive features discussed.

Dimensional analysis provides a guideline when dealing with a range of target Biot numbers. Nonetheless, real-world testing is getting complicated when the aforementioned spikiness ($\sigma$), the test duration ($t_d$), the sampling rate ($s_r$) and the noise level ($n_s$) are intertwined with the number of tick marks ($n_t$, precision-related) and the number of successful predictions ($n_{succ}$) in a Gaussian Process (GP) [1] surrogate model for data processing.

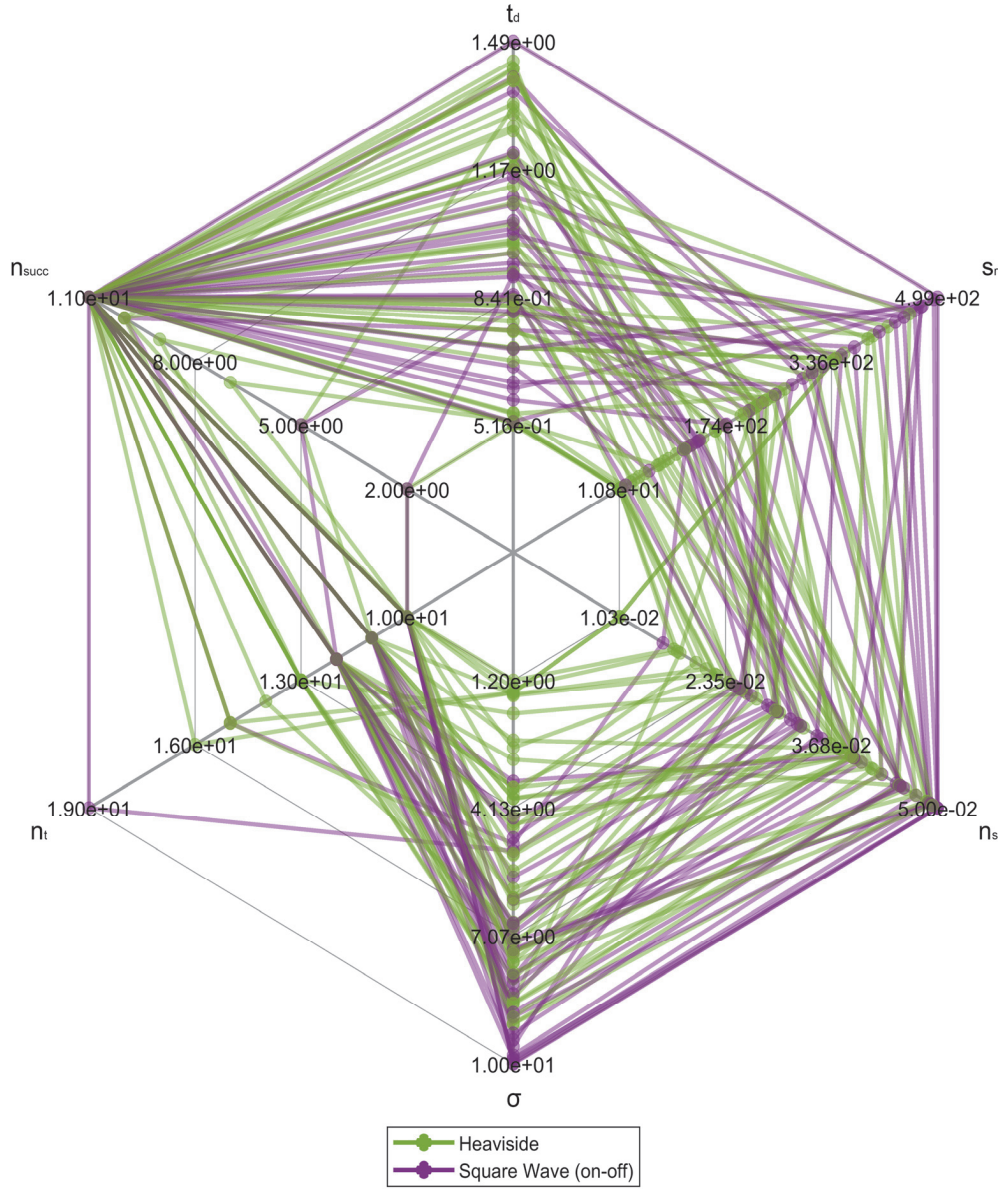

Fig. 1 - Radar chart of the Pareto front from the adapted NSGA-II [2], concerning the dimensional groups, namely, the spikiness ($\sigma$), the test duration ($t_d$), the sampling rate ($s_r$), the noise level ($n_s$), the number of tick marks ($n_t$) and the number of successful predictions ($n_{succ}$)

The lower and upper limits of the design variables ($t_d$, $s_r$ and $\sigma$) are picked following the rule of the Specific Feature of Unity (SFU) [3] in a decimals or logarithmic scale. The noise level was determined as the statistical robustness of the GP surrogate (evaluated with $n_{succ}$) and the precision of the measurement ($n_t$) is concerned, as the correlation and/or trade-off characteristics among the three are intuitively speculated aforehand.

The mode of heat perturbation, either of Heaviside or square-wave (on-off, 50% duty cycle) type, was also compared in Fig. 1. The hyperparameters of the GP surrogate were tuned via a Bayesian optimizer. A cross-validation was implemented to divide the data into $n_t$ subsets and iteratively train and validate the model on different combinations of these subsets to ensure that the model generalizes well for prediction. The simpler Heaviside receives higher preference (18.03%) over its counterpart among all cases without a single failure in prediction. Although the acquired Pareto front is indicating a higher tolerance of noise level, the rate of failures (12.86%) with the GP surrogate that comes along marks up the warning sign. The distribution for each of the variables can be identified in Fig. 1, while it is the process of acquiring a series of solutions (the Pareto front) that matters in real-world decision making which usually involves an estimation of the range of the measurement target with a few adjustable variables at disposal.

## Acknowledgments

The financial support from the National Key Research and Development Program of China (No. 2020YFB2009100), the National Natural Science Foundation of China (No. 11702045), Chongqing Municipal Bureau of Science and Technology (No. cstc2021jcyj-msxmX0004), and Chongqing Municipal Education Commission (No. KJQN202200809) is hereby acknowledged.